\documentstyle[11pt]{article} 

\newcommand{\np}[3]{{\sl Nucl. Phys.} {\bf #1} (19#2) #3}
\newcommand{\pl}[3]{{\sl Phys. Lett.} {\bf #1} (19#2) #3}
\newcommand{\pr}[3]{{\sl Phys. Rev.} {\bf #1} (19#2) #3}

\newcommand{\zp}[3]{{\sl Z. Phys.} {\bf #1} (19#2) #3}

\newcommand{\cpc}[3]{{\sl Comp. Phys. Commun.} {\bf #1} (19#2) #3}

\newcommand{\hep}[1]{{\sl hep--ph/}{#1}}

\newcommand{\phanst}{$\hphantom{\sigma^{s+t}\ }$}
\newcommand{\phanat}{$\hphantom{A_{FB}^{s+t}\ }$}
\newcommand{\phanss}{$\hphantom{\sigma^{s}\ }$}
\newcommand{\phanas}{$\hphantom{A_{FB}^{s}\ }$} 
\newcommand{\pb}{\,\mbox{pb}}
\newcommand{\pbb}{\,\mbox{\bf pb}}
\newcommand{\pe}{\,\%\:}
\newcommand{\pc}{\,\%}

\renewcommand{\thefootnote}{\fnsymbol{footnote}}

\begin{document}

\begin{titlepage}

\begin{center}
\baselineskip25pt

{\Large\sc Large-Angle Bhabha Scattering at LEP\,1}

\end{center}

\setcounter{footnote}{3}

\vspace{1cm}

\begin{center}
\baselineskip12pt

{\large\sc 
Wim Beenakker$^1$\footnote{Research supported by a fellowship of the Royal 
                           Dutch Academy of Arts and Sciences.}\vspace{0.5cm}\\
and \vspace{0.5cm}\\
Giampiero Passarino$^2$} 

\vspace{1cm}

$^1$ Instituut--Lorentz, University of Leiden, The Netherlands 
\\[5mm]
$^2$ Dipartimento di Fisica Teorica, Universit\`a di Torino, Italy \\
     INFN, Sezione di Torino, Italy

\vspace{0.3cm}

\end{center}

\vspace{2cm}

\begin{abstract}
  \normalsize \noindent A critical assessment is given of the theoretical 
  uncertainty in the predicted cross-sections for large-angle Bhabha
  scattering at LEP\,1, with or without $t$-channel subtraction. To this end a 
  detailed comparison is presented of the results obtained with the programs 
  ALIBABA and TOPAZ0. Differences in the implementation of the radiative 
  corrections and the effect of missing higher-order terms are critically 
  discussed.
\end{abstract}

\end{titlepage}

\def\thefootnote{\arabic{footnote}} \setcounter{footnote}{0}

\setcounter{page}{1}

\section{Introduction}

In 1995 the $Z$ phase of LEP has come to an end and at present the ultimate 
analysis of the data is imminent. This involves in particular the completion 
of the line-shape analysis, including the final LEP energy calibration. 
Consequently, the safest possible estimate of the theoretical accuracy is of 
the upmost importance. It should be noted that the LEP\,1 data (1990--1995) 
have been taken in the energy ($\sqrt{s}\,$) range 
$| \sqrt{s} - M_{_Z} | < 3\;$GeV and consist of the hadronic 
and leptonic cross-sections, the leptonic forward--backward asymmetries, the 
various polarization asymmetries, the partial widths, and the quark 
forward--backward asymmetries. All this makes it mandatory to assess the 
theoretical precision of the available programs for different channels and for 
energies up to $\sim 3\;$GeV away from the resonance.

In this note we focus on the electron (Bhabha) channel. Bhabha scattering is 
measured with remarkable precision at LEP\,1/SLC in two complementary 
kinematical regions: small and large scattering angles. The former plays a 
crucial role in determining the luminosity. The latter is essential for  
extracting the $Z$-boson properties. The main message of this note concerns 
an update of the theoretical precision in the large-angle regime. This 
precision depends on the beam energy and on the event selection used,
the worst case being at a few GeV above the $Z$ resonance. There the different
programs are found to deviate by as much as 1\%. This should be contrasted
with the expected experimental systematic errors \cite{precision}, displayed 
in Table~\ref{syst} (an example of the statistical errors \cite{quast}
is given in Table~\ref{stat}).
In this note we present a detailed analysis of the observed deviations and 
translate this into estimates for the theoretical precision.
\begin{table}[ht]
\begin{center}
\begin{tabular}{|c|c|c|c|c|}
\hline
& ALEPH & DELPHI & L3 & OPAL  \\
\hline\hline
$\sigma_e$ & $0.30/0.30/0.31\%$ & $0.59/0.54/0.75\%$ & $0.30/0.23/1.0\%$
& $ 0.23/0.24/-\%$  \\
$A^e_{_{FB}}$ & $0.0018/0.0019/0.0020$ & $0.0025/0.0022/0.0021$ &
$0.003/0.003/0.01$ & $0.0016/0.0016/0.002$  \\
\hline 
\end{tabular}
\end{center}
\caption[]{The (preliminary) experimental systematic errors for the years 
           1993/1994/1995 at the $Z$ peak, not including the common 
           uncertainty due to the LEP energy calibration.}
\label{syst}
\end{table}
\begin{table}[ht]
\begin{center}
\begin{tabular}{|l|c||c|c|c|}
  \hline
  1990    & $\sqrt{s}\;$ [GeV]  & 88.223   & 91.215   & 94.202   \\
  \cline{2-5}
  results & stat. error         & 3.8\%    & 2.5\%    & 3.1\%    \\
  \hline \hline
  1991    & $\sqrt{s}\;$ [GeV]  & 88.464   & 91.207   & 93.701   \\
  \cline{2-5}
  results & stat. error         & 3.3\%    & 2.2\%    & 3.1\%    \\
  \hline \hline
  1992    & $\sqrt{s}\;$ [GeV]  & $-$      & 91.270   & $-$      \\
  \cline{2-5}
  results (SICAL) & stat. error & $-$      & 1.6\%    & $-$      \\
  \hline \hline
  1993    & $\sqrt{s}\;$ [GeV]  & 89.43204 & 91.18718 & 93.01527 \\
  \cline{2-5}
  results & stat. error         & 1.2\%    & 1.6\%    & 1.1\%    \\
  \hline \hline
  1994    & $\sqrt{s}\;$ [GeV]  & $-$      & 91.19677 & $-$      \\
  \cline{2-5}
  results & stat. error         & $-$      & 0.7\%    & $-$      \\
  \hline \hline
  1995    & $\sqrt{s}\;$ [GeV]  & 89.43952 & 91.28190 & 92.96812 \\
  \cline{2-5}
  results & stat. error         & 1.2\%    & 2.2\%    & 1.0\%    \\
  \hline 
\end{tabular}
\end{center}
\caption[]{Sample of statistical errors for the electron cross-section from 
           ALEPH. For the various runs the highest and lowest energies are 
           shown, as well as the energy closest to the $Z$ peak.}
\label{stat}
\end{table}

\section{Comparison of ALIBABA and TOPAZ0}

In order to assess the theoretical uncertainties in the predictions for 
large-angle Bhabha scattering (LABS), we first perform a detailed numerical 
comparison of the programs ALIBABA \cite{ALIBABA} and TOPAZ0 \cite{TOPAZ0}. 
Previous comparisons can be found in~\cite{YR95} and also in~\cite{them}.
The input parameters chosen for the comparison are: $M_{_Z}=91.1863\;$GeV, 
$m_t=175.6\;$GeV, $M_{_H}=300\;$GeV, and $\alpha_s(M_{_Z}^2)=0.118$. The other
Standard Model parameters we take from Ref.~\cite{PDG96}. The acceptance cuts
for the final-state particles consist of a minimum energy for both particles 
($1\;$GeV), an angular acceptance for the electron 
($40^\circ < \vartheta_{e^-} < 140^\circ$), and a maximum acollinearity angle 
($10^\circ$ or $25^\circ$). For the energy we take the characteristic LEP\,1
energies: $\sqrt{s}=$ 88.45, 89.45, 90.20, 91.19, 91.30, 91.95, 93.00, and
$93.70\;$GeV.  

\begin{table}[ht]
\begin{center}
\begin{tabular}{|c||c|c|c|c|c|c|c|c|}
  \hline
  & \multicolumn{8}{c|}{LEP\,1 energy in GeV} \\
  \cline{2-9}
  & 88.45 & 89.45 & 90.20 & 91.19 & 91.30 & 91.95 & 93.00 & 93.70 \\
  \hline \hline
  \multicolumn{9}{|c|}{maximum acollinearity angle: $10^\circ$} \\
  \hline \hline
  $\sigma^{s+t}$     (T) &  457.08 & 644.86 & 912.06 & 1185.70 
                         & 1164.82 & 873.50 & 476.64 &  351.80 \\
  \phanst            (A) &  457.71 & 644.78 & 911.43 & 1184.59 
                         & 1163.71 & 876.40 & 480.23 &  355.31 \\ 
  \phanst ($\,\delta\,$) & $-0.14\pc$ & $+0.01\pc$ & $+0.07\pc$ & $+0.09\pc$
                         & $+0.10\pc$ & $-0.33\pc$ & $-0.75\pc$ & $-1.00\pc$ \\
  \hline
  $A_{FB}^{s+t}$ \,\,(T) & $+0.4448$ & $+0.3411$ & $+0.2492$ & $+0.1386$ 
                         & $+0.1298$ & $+0.1008$ & $+0.1298$ & $+0.1788$ \\
  \phanat        \,\,(A) & $+0.4454$ & $+0.3409$ & $+0.2489$ & $+0.1389$ 
                         & $+0.1301$ & $+0.1020$ & $+0.1315$ & $+0.1818$ \\
  \phanat        (T$-$A) & $-0.0006$ & $+0.0002$ & $+0.0003$ & $-0.0003$
                         & $-0.0003$ & $-0.0012$ & $-0.0017$ & $-0.0030$ \\  
  \hline \hline
  $\sigma^{s}$       (T) & 172.94 & 331.55 & 590.93 & 994.27
                         & 998.32 & 820.80 & 461.49 & 329.61 \\ 
  \phanss            (A) & 173.60 & 332.09 & 590.72 & 991.93 
                         & 996.78 & 821.13 & 463.35 & 331.49 \\ 
  \phanss ($\,\delta\,$) & $-0.38\pc$ & $-0.16\pc$ & $+0.04\pc$ & $+0.24\pc$ 
                         & $+0.15\pc$ & $-0.04\pc$ & $-0.40\pc$ & $-0.57\pc$ \\
  \hline
   $A_{FB}^{s}$  \,\,(T) & $-0.2202$ & $-0.1380$ & $-0.0761$ & $+0.0004$
                         & $+0.0081$ & $+0.0487$ & $+0.0980$ & $+0.1225$ \\
  \phanas        \,\,(A) & $-0.2209$ & $-0.1386$ & $-0.0774$ & $-0.0008$
                         & $+0.0072$ & $+0.0485$ & $+0.0977$ & $+0.1225$ \\
  \phanas        (T$-$A) & $+0.0007$ & $+0.0006$ & $+0.0013$ & $+0.0012$
                         & $+0.0009$ & $+0.0002$ & $+0.0003$ & $+0.0000$ \\ 
  \hline \hline
  \multicolumn{9}{|c|}{maximum acollinearity angle: $25^\circ$} \\
  \hline \hline
  $\sigma^{s+t}$     (T) &  485.17 & 674.89 & 945.00 & 1221.13 
                         & 1200.16 & 905.25 & 503.79 &  377.59 \\
  \phanst            (A) &  484.05 & 673.91 & 944.73 & 1220.49 
                         & 1199.51 & 907.15 & 504.73 &  378.40 \\ 
  \phanst ($\,\delta\,$) & $+0.23\pc$ & $+0.15\pc$ & $+0.03\pc$ & $+0.05\pc$
                         & $+0.05\pc$ & $-0.21\pc$ & $-0.19\pc$ & $-0.21\pc$ \\
  \hline
  $A_{FB}^{s+t}$ \,\,(T) & $+0.4605$ & $+0.3554$ & $+0.2613$ & $+0.1501$ 
                         & $+0.1417$ & $+0.1175$ & $+0.1584$ & $+0.2136$ \\
  \phanat        \,\,(A) & $+0.4576$ & $+0.3521$ & $+0.2596$ & $+0.1484$ 
                         & $+0.1402$ & $+0.1173$ & $+0.1580$ & $+0.2130$ \\
  \phanat        (T$-$A) & $+0.0029$ & $+0.0033$ & $+0.0017$ & $+0.0017$
                         & $+0.0015$ & $+0.0002$ & $+0.0004$ & $+0.0006$ \\  
  \hline \hline
  $\sigma^{s}$       (T) &  176.31 & 336.84 & 599.25 & 1007.03
                         & 1011.10 & 831.43 & 468.16 &  334.94 \\
  \phanss            (A) &  177.43 & 338.44 & 601.29 & 1008.13
                         & 1011.67 & 833.29 & 469.57 &  335.92 \\
  \phanss ($\,\delta\,$) & $-0.64\pc$ & $-0.48\pc$ & $-0.34\pc$ & $-0.11\pc$ 
                         & $-0.06\pc$ & $-0.22\pc$ & $-0.30\pc$ & $-0.29\pc$ \\
  \hline
   $A_{FB}^{s}$  \,\,(T) & $-0.2235$ & $-0.1404$ & $-0.0777$ & $-0.0004$
                         & $+0.0073$ & $+0.0480$ & $+0.0967$ & $+0.1203$ \\
  \phanas        \,\,(A) & $-0.2227$ & $-0.1406$ & $-0.0777$ & $-0.0007$
                         & $+0.0069$ & $+0.0481$ & $+0.0976$ & $+0.1210$ \\
  \phanas        (T$-$A) & $-0.0008$ & $+0.0002$ & $+0.0000$ & $+0.0003$
                         & $+0.0004$ & $-0.0001$ & $-0.0009$ & $-0.0007$ \\
  \hline 
\end{tabular}
\end{center}
\caption[]{Comparison of TOPAZ0 (T) and ALIBABA (A) for the cross-section 
           (in pb) and the forward--backward asymmetry. The TOPAZ0 results do
           not include initial-state pair production. The full Bhabha 
           results are indicated by the superscript ``$s+t$'', the s-channel 
           contributions by ``$s$''. The quantity $\delta$ stands for the 
           relative deviation $100\%\,.\,(T-A)/T$. The input parameters can
           be found in the text.}
\label{comparison}
\end{table}
In Table~\ref{comparison} we present the comparison for the cross-sections 
($\sigma$) and forward--backward asymmetries ($A_{FB}$). Besides the results 
for the full Bhabha process, we also show the pure $s$-channel contributions. 
The reason for that is twofold. First, the $s$-channel 
branch of TOPAZ0 allows a systematic inclusion of the full ${\cal O}(\alpha^2)$
QED corrections \cite{YR95}, which are only present in leading-log 
approximation in ALIBABA. In this way some of the statements that will be made 
for the full Bhabha process can be checked in the $s$-channel mode.
Secondly, the so-called $t$-channel subtraction procedure for extracting the 
pure $s$-channel part is still very popular among the experimental 
Collaborations. As such comparisons of all components
of LABS are well deserving the effort.

The numerical precision of the TOPAZ0 numbers is 0.00001 in $A_{FB}$ and 
$0.001\%$ (relative precision) in $\sigma$. For ALIBABA the numerical precision
is much less in view of the five-dimensional VEGAS integration, i.e.~0.0006
in $A_{FB}$ and 0.05\% in $\sigma$. The technical precision of the results, 
however, will not affect our conclusions. Whereas the observed LABS
differences between TOPAZ0 and ALIBABA are largely acceptable for
a maximum acollinearity angle of $25^\circ$, this is certainly not true for
$10^\circ$. In that case large deviations up to 1\% show up for energies above
the $Z$ resonance, as already discussed in~\cite{YR95}.

In Table~\ref{difference} we try to quantify where this 
difference is coming from. 
\begin{table}[ht]
\begin{center}
\begin{tabular}{|c||c|c|c|c|c|c|c|c|}
  \hline
  & \multicolumn{8}{c|}{LEP\,1 energy in GeV} \\
  \cline{2-9}
  & 88.45 & 89.45 & 90.20 & 91.19 & 91.30 & 91.95 & 93.00 & 93.70 \\
  \hline \hline
  \multicolumn{9}{|c|}{maximum acollinearity angle: $10^\circ$} \\
  \hline \hline
  $\mbox{NL}_A$ [pb]    & $- 8.05$ & $-11.54$ & $-15.45$ & $-17.80$ 
                        & $-17.52$ & $-10.48$ & $- 4.71$ & $- 2.88$ \\
  \hline
  $\Delta_{NL}$         & $-0.33\pc$ & $-0.39\pc$ & $-0.54\pc$ & $-0.64\pc$
                        & $-0.60\pc$ & $-0.62\pc$ & $-0.42\pc$ & $-0.44\pc$ \\
  $\Delta_{FSR}$        & $+0.08\pc$ & $+0.11\pc$ & $+0.02\pc$ & $-0.17\pc$
                        & $-0.17\pc$ & $-0.48\pc$ & $-0.69\pc$ & $-0.86\pc$ \\
  $\delta-\Delta_{FSR}$ & $-0.22\pc$ & $-0.10\pc$ & $+0.05\pc$ & $+0.27\pc$
                        & $+0.26\pc$ & $+0.14\pc$ & $-0.06\pc$ & $-0.14\pc$ \\
  \hline \hline
  \multicolumn{9}{|c|}{maximum acollinearity angle: $25^\circ$} \\
  \hline \hline
  $\mbox{NL}_A$ [pb]    & $-4.14$ & $-4.59$ & $-4.17$ & $-3.74$ 
                        & $-3.54$ & $-1.48$ & $-1.89$ & $-1.49$ \\
  \hline
  $\Delta_{NL}$         & $+0.23\pc$ & $+0.02\pc$ & $-0.24\pc$ & $-0.35\pc$
                        & $-0.35\pc$ & $-0.39\pc$ & $-0.05\pc$ & $+0.02\pc$ \\
  $\Delta_{FSR}$        & $+0.34\pc$ & $+0.16\pc$ & $-0.08\pc$ & $-0.22\pc$
                        & $-0.23\pc$ & $-0.36\pc$ & $-0.15\pc$ & $-0.13\pc$ \\
  $\delta-\Delta_{FSR}$ & $-0.11\pc$ & $-0.01\pc$ & $+0.11\pc$ & $+0.27\pc$
                        & $+0.28\pc$ & $+0.15\pc$ & $-0.04\pc$ & $-0.09\pc$ \\
  \hline 
\end{tabular}
\end{center}
\caption[]{An analysis of the differences between TOPAZ0 and ALIBABA for 
           large-angle Bhabha scattering. The various entries are explained
           in the text.}
\label{difference}
\end{table}
In this context it is important to note that the
main conceptual difference between TOPAZ0 and ALIBABA in the Bhabha channel 
lies in the implementation of the non-leading-log QED corrections. Both 
programs are based on the structure-function method for calculating the 
(dominant) leading-log corrections [$\propto (\alpha L/\pi)^n$, with $n=1,2$ 
and $L=\log(s/m_e^2)$]. When it comes to the non-leading-log corrections both
programs use quite different approaches. TOPAZ0 was designed to be an 
efficient {\it fitter} of realistic observables around the $Z$ resonance, 
where LABS is effectively dominated by the annihilation channel.
The structure functions in TOPAZ0 are based on the iterative solution of
the master evolution equation, which accounts for the well-known second-order
electron form factor, and therefore they reproduce the sub-leading terms
[$\propto (\alpha/\pi)^n L^m$, with $n=1,2$ and $n>m$] for the $s$ channel.%
\footnote{In a realistic (i.e.~not fully-extrapolated) set-up some 
          ${\cal O}(\alpha)$ hard-photonic non-log corrections are missing in 
          the $s$ channel. However, their effect is tiny around the $Z$ 
          resonance \cite{pavians}.}
This procedure, however, does not reproduce the (unknown) correct answer
for the sub-leading terms in the $t$ channel.

In TOPAZ0 a part of the sub-leading terms in the $t$ channel enters through a 
final-state-radiation factor, which is subsequently convoluted with the 
structure functions. In this way part of the ${\cal O}(\alpha)$ non-log terms 
are implemented, as well as a subset of sub-leading second-order terms 
$\propto \alpha^2 L/\pi^2$. These sub-leading higher-order terms are not 
present in ALIBABA. However, in the ALIBABA approach the full set of non-log 
${\cal O}(\alpha)$ corrections are determined by means of the five-dimensional
VEGAS integration, needed for handling the radiative process. These non-log
corrections are listed in the $\mbox{NL}_A$ entries in Table~\ref{difference}.
The entries $\Delta_{NL}$ in Table~\ref{difference} quantify the deviations 
caused by considering non-leading terms only from final-state radiation. In 
conclusion, the main difference in the treatment of the QED corrections in both
programs can be estimated by $\,C_{F}\,(\sigma^{s+t}-\mbox{NL}_A)-\mbox{NL}_A$.
Here the ${\cal O}(\alpha)$ final-state-radiation factor $C_{F}$ is given by 
$-0.0165$ ($-0.0052$) for a maximum acollinearity angle of $10^\circ$ 
($25^\circ$). In Table~\ref{difference} we present this estimate in the form 
of a relative correction factor ($\Delta_{FSR}$) with respect to the full 
Bhabha cross-section. Comparing the estimate with the actual deviation $\delta$
in Table~\ref{comparison}, a systematic shift ($\delta-\Delta_{FSR}$) is 
observed that is roughly 
independent of the maximum acollinearity angle. The resulting shift can be 
attributed to the fact that the ALIBABA weak library is not up to date. 
This is supported by a direct comparison with the Bhabha branch of the 
program ZFITTER \cite{ZFITTER}. Another thing to note is that for a 
maximum acollinearity angle of $10^\circ$ both TOPAZ0 and ALIBABA seem to miss 
terms of the order of 0.3--0.6\% (cf. $\Delta_{NL}$ and 
$\Delta_{NL}-\Delta_{FSR}$ in Table~\ref{difference}). 
Below and near the $Z$ resonance both effects go 
in the same direction, compensating each other in the difference. This leads to
a somewhat misleading agreement between the programs. Above the 
resonance the effects have different signs, leading to an enhancement of the 
deviations. The relative deviation in the non $s$-channel components 
($\sigma^{s+t}-\sigma^s$), which is practically negligible below the 
resonance, grows up to $\sim 10\%$ on the high-energy side. 
There, however, the non $s$-channel cross-sections are rapidly decreasing 
(e.g.~roughly $23\,$pb at $\sqrt{s}=93.70\;$GeV, i.e.~$6.5\%$ of full LABS, 
versus $284\,$pb at $\sqrt{s}=88.45\;$GeV, i.e.~$62\%$ of full LABS).

\section{Theoretical error estimates for large-angle Bhabha scattering}

Having established the main sources of the differences between TOPAZ0 and 
ALIBABA,
we can now address the question of the error in the theoretical predictions
for large-angle Bhabha scattering. In the experimental analyses either
the full Bhabha cross-section is used or merely the $s$-channel contributions.
The latter are obtained through $t$-channel subtraction, i.e.~by subtracting 
the non $s$-channel contributions that involve $t$-channel gauge-boson 
exchange. The subtraction procedure is aimed at reducing the full LABS to a 
simple annihilation process to be subsequently analyzed by some up-to-date
(from the point of view of the WEAK/QCD library) program. 

Correspondingly we will present the theoretical errors
for both procedures. The bulk of the theoretical errors are due to missing
QED corrections: missing non-log ${\cal O}(\alpha)$ corrections in
TOPAZ0, missing higher-order sub-leading effects from the structure-function 
convolution of the non-log ${\cal O}(\alpha)$ corrections in ALIBABA,
missing effects in both programs from the (unknown) sub-leading terms in the 
structure functions, and missing initial-state pair-production effects in 
ALIBABA. In addition there is the usual calorimetric measurement problem,
which arises if the cross-section
is not inclusive in the energy of the outgoing fermions.
While with the present cuts the effect is estimated to be negligible by
TOPAZ0 ($\ll 0.1\%$), for high-energy thresholds the contribution will
grow considerably. 

In general one should remember that an absolute test of the QED corrections
does not exist. QED corrections are convoluted with -- in principle --
different kernel cross-sections, so that the shift in the absolute
QED corrections also depends on the differences in the non-QED parts.

Before coming to all these QED effects, we first estimate the uncertainty in 
the weak sector. From a detailed study with TOPAZ0 it follows that different 
weak options influence the cross-sections by at most 0.06\% and the 
forward--backward asymmetry by at most 0.0002. 
In this respect the quality of the estimate has not changed since the work 
done in~\cite{YR95}, since the upgrading in TOPAZ0 and ZFITTER have been 
constantly cross-checked. The uncertainty in ALIBABA roughly equals the 
$\Delta_{FSR}-\delta$ shifts displayed in Table~\ref{difference}: 
varying $M_{_H}$ between 60 and $1000\;$GeV, and $M_{_Z}$, $m_t$, and 
$\alpha_s(M_{_Z}^2)$ within one experimental standard deviation, leads
to variations of the cross-sections and asymmetries that closely resemble the 
ones observed with TOPAZ0. The ALIBABA variations have the same sign as the 
TOPAZ0 ones and do not differ by more than 0.06\% for the cross-section and 
0.0006 for the forward--backward asymmetry, which is of the same size as the 
numerical error in the ALIBABA results. 

In the context of the QED errors, the ALIBABA error owing to the absence of
initial-state pair-production (ISPP) effects can be estimated in a 
straightforward way with the help of TOPAZ0. For the eight LEP\,1 energy 
points 
we find for the full LABS cross-section: $-0.23\%$, $-0.25\%$, $-0.26\%$, 
$-0.25\%$, $-0.24\%$, $-0.19\%$, $-0.04\%$, $+0.07\%$. 
Here the final-state $e^+e^-$ pair is required to have an invariant mass that
exceeds 50\% of the total energy $\sqrt{s}$ (i.e.~$z_{\rm min}=0.25$).
It should be noted that 
among all radiative corrections the ISPP appears as the most questionable. 
While relatively safe around the resonance, different implementations of 
ISPP start to register some disagreement at higher energies~\cite{YRLEP2} and 
strongly depend on kinematical cuts, in other words on the precise separation 
between 2-fermion and 4-fermion physics.

The missing sub-leading higher-order terms from the convolution of the non-log 
${\cal O}(\alpha)$ corrections (MSL1) in ALIBABA is harder to estimate. Here 
we deploy two methods. As can be read off from the $\mbox{NL}_A$ entries in 
Table~\ref{difference}, the non-log ${\cal O}(\alpha)$ corrections in ALIBABA 
appear in two different shapes. For a maximum acollinearity angle of 
$10^\circ$ the $\mbox{NL}_A$ contribution exhibits a resonance-like behaviour, 
with a pronounced peak around $\sqrt{s}=M_{_Z}$. In that case the leading-log 
corrections to this contribution should closely resemble the leading-log 
corrections to the lowest-order cross-section. A natural way of estimating 
these higher-order effects is by dividing the $\mbox{NL}_A$ contributions by 
the lowest-order cross-sections (in `dressed-Born' form), taking the maximum 
value of this ratio, and subsequently multiplying the leading-log corrections 
in ALIBABA by this number. Note that in principle the sign of the missing 
terms will be fixed. For a maximum 
acollinearity angle of $25^\circ$ the $\mbox{NL}_A$ contribution is rather 
flat, with a sudden jump just above the $Z$ resonance. In that case the best
way to estimate the higher-order effects is by simply multiplying the maximum
value for $|\mbox{NL}_A|$ by the factor $(4\alpha L/\pi)$. Here an 
additional factor of two is added for safety. The sign of the missing terms is
only determined for the three highest energy points, i.e.~the ones after the 
jump. Since the structure-function convolution probes the variation in the 
$\mbox{NL}_A$ contributions at lower energies, a negative sign is expected for
the higher-order corrections in these three energy points. 
Note that this second method also applies to the non $s$-channel 
cross-sections needed for the $t$-channel subtraction, irrespective of the 
maximum acollinearity angle. 
Finally, the effect from the (unknown) sub-leading terms in the structure 
functions (MSL2) can be estimated by multiplying the leading-log corrections 
by the factor $(4\alpha/\pi)$. Again a factor of two is added for safety. 
Note that the sign of the MSL2 terms is not fixed.

In Table~\ref{errorA} we summarize the estimates for the theoretical errors of
the ALIBABA program.%
\footnote{We have also derived the non $s$-channel errors for the ALEPH 
          angular acceptance, involving a $20^\circ$ maximum acollinearity 
          angle and $45^\circ < \angle(e^-,e^-_{\rm beam}) < 155^\circ$. 
          The errors read: $1.1\pb$, $1.2\pb$, $1.2\pb$, $1.1\pb$, $1.1\pb$,
          $1.2\pb$, $1.1\pb$, and $1.0\pb$. These errors are smaller than the
          ones in Table~\protect\ref{errorA}, since also the non $s$-channel 
          cross-sections themselves are smaller. The errors on the full LABS
          cross-section are given by: 0.5\%, 0.4\%, 0.4\%, 0.5\%, 0.5\%, 0.4\%,
          0.4\%, and 0.5\%.}     
In the total error the various estimates are added in quadrature. 
\begin{table}[ht]
\begin{center}
\begin{tabular}{|c||c|c|c|c|c|c|c|c|}
  \hline
  & \multicolumn{8}{c|}{LEP\,1 energy in GeV} \\
  \cline{2-9}
  & 88.45 & 89.45 & 90.20 & 91.19 & 91.30 & 91.95 & 93.00 & 93.70 \\
  \hline \hline
  \multicolumn{9}{|c|}{maximum acollinearity angle: $10^\circ$} \\
  \hline \hline
  weak     & $0.17\pe$ & $0.06\pe$ & $0.08\pe$ & $0.27\pe$
           & $0.27\pe$ & $0.15\pe$ & $0.05\pe$ & $0.12\pe$ \\
           & $0.57\pb$ & $0.63\pb$ & $0.64\pb$ & $0.39\pb$
           & $0.33\pb$ & $0.33\pb$ & $0.33\pb$ & $0.33\pb$ \\
  \hline
  pairs    & $0.23\pe$ & $0.25\pe$ & $0.26\pe$ & $0.25\pe$
           & $0.24\pe$ & $0.19\pe$ & $0.04\pe$ & $0.07\pe$ \\
           & $0.65\pb$ & $0.78\pb$ & $0.83\pb$ & $0.48\pb$
           & $0.40\pb$ & $0.32\pb$ & $0.07\pb$ & $0.12\pb$ \\
  \hline
  MSL1     & $0.34\pe$ & $0.42\pe$ & $0.47\pe$ & $0.39\pe$
           & $0.36\pe$ & $0.13\pe$ & $0.22\pe$ & $0.35\pe$ \\
           & $1.36\pb$ & $1.36\pb$ & $1.36\pb$ & $1.36\pb$
           & $1.36\pb$ & $1.36\pb$ & $1.36\pb$ & $1.36\pb$ \\
  \hline
  MSL2     & $0.23\pe$ & $0.28\pe$ & $0.32\pe$ & $0.26\pe$
           & $0.24\pe$ & $0.24\pe$ & $0.24\pe$ & $0.24\pe$ \\
           & $0.46\pb$ & $0.54\pb$ & $0.50\pb$ & $0.31\pb$
           & $0.42\pb$ & $0.76\pb$ & $0.52\pb$ & $0.35\pb$ \\
  \hline \hline
  {\bf total} & \boldmath{$0.5\pe$}  & \boldmath{$0.6\pe$} 
              & \boldmath{$0.6\pe$}  & \boldmath{$0.6\pe$}
              & \boldmath{$0.6\pe$}  & \boldmath{$0.4\pe$} 
              & \boldmath{$0.3\pe$}  & \boldmath{$0.4\pe$}  \\
              & \boldmath{$1.7\pbb$} & \boldmath{$1.8\pbb$}
              & \boldmath{$1.8\pbb$} & \boldmath{$1.5\pbb$} 
              & \boldmath{$1.5\pbb$} & \boldmath{$1.6\pbb$} 
              & \boldmath{$1.5\pbb$} & \boldmath{$1.4\pbb$} \\
  \hline \hline
  \multicolumn{9}{|c|}{maximum acollinearity angle: $25^\circ$} \\
  \hline \hline
  weak     & $0.17\pe$ & $0.06\pe$ & $0.08\pe$ & $0.27\pe$
           & $0.27\pe$ & $0.15\pe$ & $0.05\pe$ & $0.12\pe$ \\
           & $0.61\pb$ & $0.67\pb$ & $0.69\pb$ & $0.42\pb$
           & $0.38\pb$ & $0.38\pb$ & $0.38\pb$ & $0.38\pb$ \\
  \hline  
  pairs    & $0.23\pe$ & $0.25\pe$ & $0.26\pe$ & $0.25\pe$
           & $0.24\pe$ & $0.19\pe$ & $0.04\pe$ & $0.07\pe$ \\
           & $0.71\pb$ & $0.84\pb$ & $0.89\pb$ & $0.53\pb$
           & $0.45\pb$ & $0.36\pb$ & $0.08\pb$ & $0.13\pb$ \\
  \hline
  MSL1     & $0.21\pe$ & $0.15\pe$ & $0.11\pe$ & $0.08\pe$
           & $0.09\pe$ & $0.11\pe$ & $0.20\pe$ & $0.27\pe$ \\
           & $0.97\pb$ & $0.97\pb$ & $0.97\pb$ & $0.97\pb$ 
           & $0.97\pb$ & $0.97\pb$ & $0.97\pb$ & $0.97\pb$ \\
  \hline
  MSL2     & $0.17\pe$ & $0.24\pe$ & $0.28\pe$ & $0.24\pe$
           & $0.22\pe$ & $0.22\pe$ & $0.22\pe$ & $0.27\pe$ \\
           & $0.26\pb$ & $0.35\pb$ & $0.30\pb$ & $0.50\pb$ 
           & $0.62\pb$ & $0.95\pb$ & $0.71\pb$ & $0.53\pb$ \\
  \hline \hline
  {\bf total} & \boldmath{$0.4\pe$}  & \boldmath{$0.4\pe$} 
              & \boldmath{$0.4\pe$}  & \boldmath{$0.4\pe$}
              & \boldmath{$0.4\pe$}  & \boldmath{$0.3\pe$} 
              & \boldmath{$0.3\pe$}  & \boldmath{$0.4\pe$}  \\
              & \boldmath{$1.4\pbb$} & \boldmath{$1.5\pbb$} 
              & \boldmath{$1.5\pbb$} & \boldmath{$1.3\pbb$} 
              & \boldmath{$1.3\pbb$} & \boldmath{$1.5\pbb$} 
              & \boldmath{$1.3\pbb$} & \boldmath{$1.2\pbb$} \\
  \hline 
\end{tabular}
\end{center}
\caption[]{Error estimates for the large-angle Bhabha cross-section as 
           predicted with ALIBABA. The numbers in the first row of every entry
           correspond to the unsubtracted cross-sections and are given 
           relative to $\sigma^{s+t}$ (see Table~\protect\ref{comparison}). 
           The numbers in the second row (given in pb) correspond to the non 
           $s$-channel components needed for the $t$-channel subtraction.}
\label{errorA} 
\end{table}
For the unsubtracted cross-section we take the average of the two 
$\delta-\Delta_{FSR}$ shifts, displayed in Table~\ref{difference}, as a 
measure of the error in the weak corrections. For the non $s$-channel 
cross-section a constant uncertainty of 0.2\% is assumed.  
At this point one should bear in mind that the non $s$-channel cross-sections
exhibit a strong cancellation between the $t$-channel and $s$--$t$ interference
contributions at the three highest energy points (see Table~\ref{comparison}).
In order to avoid underestimating the weak and ISPP errors at these energies,
we therefore use the cross-section at $91.30\;$GeV in the error estimates.
The error estimates for the non $s$-channel cross-sections are given in pb,
rather than relative to the full $\sigma^{s+t}-\sigma^s$ results. In this way
the errors are independent of the amount of cancellation between the 
$t$-channel and $s$--$t$ interference contributions.
Note that the non $s$-channel errors are not
derived from the errors for the full Bhabha cross-section. They are derived
directly from $\sigma^{s+t}-\sigma^s$, eliminating in this way the 
correlation between the errors in $\sigma^{s+t}$ and $\sigma^s$.  

As a cross-check of the above error-estimate procedure, we have also derived
the pure $s$-channel errors. For a maximum acollinearity angle of $10^\circ$ we
find 0.4\%, 0.4\%, 0.4\%, 0.5\%, 0.5\%, 0.3\%, 0.4\%, and 0.5\%; for $25^\circ$
the errors are 0.7\%, 0.6\%, 0.5\%, 0.3\%, 0.3\%, 0.4\%, 0.4\%, and 0.4\%.
This is in good agreement with the observed $s$-channel deviations in
Table~\ref{comparison}.

The main approximation for LABS in TOPAZ0 is the leading-logarithmic one in 
the non $s$-channel components. The QED theoretical
error in TOPAZ0 is therefore growing when enlarging the angular acceptance to
smaller angles, or in general in all situations where the non $s$-channel
terms are increased.

In order to give an estimate for the theoretical error in TOPAZ0 we proceed in 
the following -- conservative -- way. First of all, we assign an overall
$0.06\%$ error to the weak corrections. This is not the consequence of a 
mistreatment of the corrections, but a safe upper bound on the uncertainty 
arising from different implementation schemes.

Next, the ISPP as implemented in TOPAZ0 is strictly speaking only valid for 
$s$-channel processes. If $\Delta_p$ is the relative ISPP effect, then 
$\Delta_p\,\sigma^s$ is the correct $s$-channel contribution. The missing 
terms are proportional to the pure $t$-channel and $s$--$t$ interference 
components. However, we refrain from assigning to LABS an overall uncertainty 
of $\pm \Delta_p\,(\sigma^{s+t}-\sigma^{s})$, since the difference 
$\sigma^{s+t}-\sigma^s$ becomes very small above the peak, while 
the individual $t$-channel and $s$--$t$ interference terms can be as large as 
below the peak. Clearly this cancellation cannot be transferred to the error 
estimates. We use two different procedures. First we proceed by assigning to 
the cross-section an error $\pm \Delta_p\,\sigma_{\rm max}$, where
$\sigma_{\rm max} = \max\{\sigma^s, \sigma^{s+t}-\sigma^s\}$. The obtained
error is roughly half of $\Delta_p\,\sigma^{s+t}$ below the resonance and 
almost coincides with $\Delta_p\,\sigma^{s+t}$ above it. 
Alternatively we compute $\sigma_{\rm abs}$, the sum of the $t$-channel term 
and the absolute value of the $s$--$t$ interference, and assign
an error $\pm \Delta_p\,\sigma_{\rm abs}$.
In the two procedures the errors read: $0.14\%$, $0.13\%$, $0.17\%$, $0.21\%$, 
$0.21\%$, $0.18\%$, $0.04\%$, $0.07\%$ and $0.17\%$, $0.14\%$, $0.11\%$,
$0.05\%$, $0.05\%$, $0.08\%$, $0.03\%$, $0.07\%$. There are appreciable
differences only around the peak. In the summary we will report the average 
of the two procedures.

Similarly the MSL2 effect in TOPAZ0 is only active for the $t$-channel and 
$s$--$t$ interference contributions. For the associated uncertainty we take 
again the average of MSL2\,$\sigma_{\rm max}$ and MSL2\,$\sigma_{\rm abs}$, 
with MSL2 as estimated by ALIBABA. We have also performed a consistency check 
by computing the TOPAZ0 non $s$-channel cross-sections with the change 
$L \to ( 1 \pm 4\,\frac{\alpha}{\pi})\,L$ 
in the definition of $\beta$, appearing in the structure functions.
The resulting effect is roughly equal to the one obtained in the 
$\sigma_{\rm abs}$ method. Nevertheless we prefer to be more conservative and 
to average with the error from the $\sigma_{\rm max}$ method.

Another source of theoretical error remains in the missing non-log 
${\cal O}(\alpha)$ corrections (MNL) from QED radiation.
TOPAZ0 has some of these non-log terms, which have been included into the 
structure functions through the exact second-order $s$-channel vertex. 
We have computed this contribution, $\Delta^T_{NL}$, and found that it is 
largely due to the $\pi^2/3-2$ term in the K-factor. Next we assume that 
$-\Delta_{NL}$, as taken from ALIBABA, is the exact contribution and we 
estimate the uncertainty to be $-\Delta_{NL}-\Delta^T_{NL}$.

In Table~\ref{errorT} we summarize the estimates for the theoretical errors of
the TOPAZ0 program. In the total error the various estimates are again 
added in quadrature.
\begin{table}[ht]
\begin{center}
\begin{tabular}{|c||c|c|c|c|c|c|c|c|}
  \hline
  & \multicolumn{8}{c|}{LEP\,1 energy in GeV} \\
  \cline{2-9}
  & 88.45 & 89.45 & 90.20 & 91.19 & 91.30 & 91.95 & 93.00 & 93.70 \\
  \hline \hline
  \multicolumn{9}{|c|}{maximum acollinearity angle: $10^\circ$} \\
  \hline \hline
  weak          & $0.06\pc$ & $0.06\pc$ & $0.06\pc$ & $0.06\pc$
                & $0.06\pc$ & $0.06\pc$ & $0.06\pc$ & $0.06\pc$ \\
  \hline
  pairs         & $0.16\pc$ & $0.14\pc$ & $0.14\pc$ & $0.13\pc$
                & $0.13\pc$ & $0.13\pc$ & $0.04\pc$ & $0.07\pc$ \\
  \hline
  MSL2          & $0.16\pc$ & $0.15\pc$ & $0.17\pc$ & $0.14\pc$
                & $0.13\pc$ & $0.17\pc$ & $0.21\pc$ & $0.23\pc$ \\
  \hline
  MNL           & $0.00\pc$ & $0.07\pc$ & $0.23\pc$ & $0.33\pc$
                & $0.29\pc$ & $0.28\pc$ & $0.03\pc$ & $0.01\pc$ \\
  \hline \hline
  {\bf total}   & \boldmath{$0.2\pc$} & \boldmath{$0.2\pc$} 
                & \boldmath{$0.3\pc$} & \boldmath{$0.4\pc$}
                & \boldmath{$0.4\pc$} & \boldmath{$0.4\pc$}
                & \boldmath{$0.2\pc$} & \boldmath{$0.3\pc$} \\
  \hline \hline
  \multicolumn{9}{|c|}{maximum acollinearity angle: $25^\circ$} \\
  \hline \hline
  weak          & $0.06\pc$ & $0.06\pc$ & $0.06\pc$ & $0.06\pc$
                & $0.06\pc$ & $0.06\pc$ & $0.06\pc$ & $0.06\pc$ \\
  \hline  
  pairs         & $0.17\pc$ & $0.14\pc$ & $0.14\pc$ & $0.14\pc$
                & $0.13\pc$ & $0.13\pc$ & $0.04\pc$ & $0.07\pc$ \\
  \hline
  MSL2          & $0.12\pc$ & $0.13\pc$ & $0.15\pc$ & $0.13\pc$
                & $0.12\pc$ & $0.15\pc$ & $0.19\pc$ & $0.25\pc$ \\
  \hline
  MNL           & $0.65\pc$ & $0.41\pc$ & $0.12\pc$ & $0.00\pc$
                & $0.00\pc$ & $0.01\pc$ & $0.42\pc$ & $0.56\pc$ \\
  \hline \hline
  {\bf total}   & \boldmath{$0.7\pc$} & \boldmath{$0.5\pc$} 
                & \boldmath{$0.3\pc$} & \boldmath{$0.2\pc$}
                & \boldmath{$0.2\pc$} & \boldmath{$0.2\pc$} 
                & \boldmath{$0.5\pc$} & \boldmath{$0.6\pc$} \\
  \hline 
\end{tabular}
\end{center}
\caption[]{Error estimates for the large-angle Bhabha cross-section as 
           predicted with TOPAZ0. }
\label{errorT} 
\end{table}

Finally we come to the theoretical errors on the forward--backward asymmetry.
As can be seen from Table~\ref{comparison},
the forward--backward asymmetry $A_{FB}^{s+t}$ shows the typical $t$-channel 
effect, which makes it unique among leptonic asymmetries. At the peak we 
observe a T$-$A absolute difference of $-0.0003$ ($+0.0017$) for an average 
asymmetry of $+0.1388$ ($+0.1493$) at 
$\theta_{\rm acoll} = 10^\circ\ (25^\circ)$. This deteriorates up to a 
$-0.0030$ ($+0.0033$) difference for an average asymmetry of 
$+0.1803$ ($+0.3538$) at $\sqrt{s} =  93.70\;$GeV ($89.45\;$GeV) and 
$\theta_{\rm acoll} = 10^\circ\ (25^\circ)$. 
The $s$-channel asymmetry, $A_{FB}^s$, is instead very small in the $Z$ peak 
region. For this quantity the disagreement is globally contained within an 
absolute difference of $0.0013$. This is no surprise, since it has been shown 
in~\cite{YR95} that differences among programs are weak-dominated around the 
peak and QED-dominated far from the peak only.
For $A_{FB}^{s+t,s}$ we assign to both codes an error $\pm \Delta_{_{FB}}$,
with $\Delta_{_{FB}}$ given by the half-difference of the predictions 
(i.e.~half of the T$-$A row of table~\ref{comparison}).

\section{Conclusions}

The main emphasis of this note was on presenting the safest possible estimate 
for the theoretical accuracy of the large-angle Bhabha scattering calculations.
As usual, in estimating theoretical uncertainties one proceeds by comparing
the results of different programs starting from a common set of input
parameters and kinematical cuts. In our case we have used the predictions
of ALIBABA and TOPAZ0. The registered differences give a rough idea
of the uncertainty associated with different implementations of radiative
corrections. Successively one tries to estimate the internal accuracy of
the programs by deriving bounds on the effects of missing parts of the 
calculations.

Our analysis shows a substantial agreement between the two programs for larger 
acollinearity cuts, with a deterioration towards small acollinearity cuts and 
energies on the high-energy side of the resonance. For a maximum acollinearity
angle of $10^\circ$ we find deviations of $-0.14\%,\,+0.09\%,\,-1.00\%$ at 
$M_{_Z}-2.5\;$GeV, $M_{_Z}$, $M_{_Z}+2.5\;$GeV
for the full Bhabha cross-section ($-0.38\%,\,+0.24\%,\,-0.57\%$ for the 
$s$ channel alone). The main sources of these deviations seem understood.

As for the internal estimate, we find that for the tightest acollinearity cut
both ALIBABA and TOPAZ0 miss terms of the order of 0.2--0.6\%, depending on the
c.m.s energy. For TOPAZ0 the bulk of the effect is in the non $s$-channel.
From the point of view of TOPAZ0 the analysis of the full Bhabha cross-section
is anyhow preferable to the $t$-channel subtraction procedure, provided
that the $s$ channel dominates. The size of the uncertainty in ALIBABA 
resembles the one in TOPAZ0, although this is somehow accidental in view of the
different origins of the contributing effects. The ALIBABA program is better
suited for the $t$-channel subtraction procedure, with an uncertainty in the
non $s$-channel components ranging from $1.0\pb$ to $1.8\pb$ for the various
c.m.s. energies and angular acceptances.
If subtraction is performed then the TOPAZ0 program is better suited for the 
$s$-channel analysis due to the high precision achieved there.


\begin{thebibliography}{99}

\bibitem{precision}
D.~Ward, talk presented at the EPS-HEP-97 conference, Jerusalem, 
August 19--26, 1997, to appear in the proceedings (see also internal note
LEPEWWG/97-02 of the LEP Electroweak Working Group).

\bibitem{quast}
G.~Quast, private communication.

\bibitem{ALIBABA}
W.~Beenakker, F.A.~Berends and S.C.~van der Marck, \np{B349}{91}{323} and
\pl{B251}{90}{299}.

\bibitem{TOPAZ0}
G.~Montagna et al., \np{B401}{93}{3} and \cpc{76}{93}{328}.

\bibitem{YR95}
D.~Bardin et al., in {\it Reports of the working group on precision 
calculations for the Z resonance\/}, eds.~D.~Bardin, W.~Hollik and
G.~Passarino, (CERN-95-03, Geneva, 1995) p.\,7 and \hep{9709229}.

\bibitem{them} 
S.~Jadach et al., \pl{B390}{97}{298}.

\bibitem{PDG96}
Particle Data Group, R.M.~Barnett et al., \pr{D54}{96}{1}.

\bibitem{pavians} 
G.~Montagna, O.~Nicrosini and F.~Piccinini, \zp{C76}{97}{45}.

\bibitem{ZFITTER} 
D.~Bardin et al., CERN-TH-6443 and \hep{9412201}.

\bibitem{YRLEP2} 
{\it Standard Model Processes\/}, F.~Boudjema et al., 
in {\it Physics at LEP2\/}, eds.~G.~Altarelli et al., 
(CERN-96-01, Geneva, 1996) Vol.\,1, p.\,207 and \hep{9601224}.

\end{thebibliography}
\end{document}